# Lemons and Learning: Adult and Juvenile Free-Ranging Dogs Navigate Aversive Foraging Challenges Differently

Tuhin Subhra Pal [1], Prithwiraj Debnath [2†], Sagarika Biswas [1†], Anindita Bhadra [1*]

[1] Department of Biological Sciences, Indian Institute of Science Education and Research, Kolkata

[2] Department of Biological Sciences, Indian Institute of Science Education and Research, Berhampur

*Corresponding Author

Anindita Bhadra (abhadra@iiserkol.ac.in)

**Abstract**

Understanding how animals make foraging decisions in challenging or unpleasant contexts sheds light on the processes that underlie cognitive development and the evolution of adaptive foraging techniques in complex ecological settings. In order to investigate how age, gender, and environmental cues impact decision-making and scavenging behaviour, this study explores how juvenile and adult free-ranging dogs (FRDs) respond to aversive stimuli, specifically different concentrations of lemon juice. We conducted a three-bowl choice experiment using lemon juice at concentrations of 25%, 33.3%, and 50% with 73 juvenile free-ranging dogs (FRDs). To evaluate developmental shifts in foraging behaviour, juvenile FRD data were compared with adult behavioural data from a prior experiment conducted with the same protocol. The findings indicated that juvenile dogs' foraging behaviour was consistent under provided circumstances, suggesting that their capacity to adapt to aversive acidic stimuli was limited, possibly due to their inexperience and developing cognitive abilities. Adult dogs, on the other hand, exhibited selective foraging by preferring lower acidity and demonstrating more deliberate foraging strategies. Age, sex, and acidity all had significant impact on eating behaviour, with males displaying a higher tolerance to unpleasant stimuli. Markov chain analysis showed repetitive decision-making patterns, especially in adults, indicating continuous sensory evaluation before food consumption, while strategy-making behaviour increased the likelihood of food consumption in juveniles, suggesting juveniles are possibly beginning to build adaptive skills. These findings demonstrate the dynamic interaction of developmental stage, sensory processing, and individual diversity, in developing of foraging strategies of FRDs. This offers new perspectives to understand the cognitive basis of scavenging behaviour in aversive ecological conditions like urban and human-dominated landscapes.

**Keywords**

Free-ranging dogs, Avoidance, Food preference, Juvenile, Lemon

## 1. Introduction

Food preferences of animals have strong ecological underpinnings and can influence their behaviour to a large extent. Some species are generalists when it comes to food, while others are specialists with particular preferences. According to Losos & Ricklefs (2009), feeding habits can result in adaptations and speciation, as observed in Galapagos finches. Local factors such as taste, flavour, and visual appeal influence food preference, determining a food item's palatability (Cole & Endler, 2015; Provenza, 1995). Palatability and the hedonic value of food, as well as the type and quantity consumed, are influenced by flavour, which is primarily determined by taste and smell (Bellisle, 1989; Baumont, 1996; Ishii et al., 2003).

Animals exhibit food preferences through intake ratios and consumption behaviour, with diet selection serving as a measure of preference (Forbes & Kyriazakis, 1995). Taste order and surface characteristics impact food palatability (Kenney & Black, 1984), while dietary exposure can modify palatability (Chapple & Lynch, 1986; Provenza & Balph, 1987). Changes in habitat affect food availability, influencing species survival. Animals evaluate what, where, and how long to forage (Stephens, 2008). Optimal foraging theory suggests that fluctuating food accessibility in nature requires variation (Pyke et al., 1977). Urbanization and habitat loss expose animals to altered environments, often reducing species diversity (Charnov & Orians, 2006). However, some animals exhibit remarkable adaptations, providing insight into survival mechanisms. In the Global South, free-ranging dogs (FRDs) serve as excellent models for examining urban adaptation (Biswas et al., 2023).

Early experiences during critical developmental stages are crucial in shaping an animal's social, sexual, and feeding behaviours (Fuller and Waller, 1962; Scott, 1962; Hess, 1962). Juveniles, initially inexperienced with complex skills (Custance et al., 1999), are prone to more foraging mistakes. However, driven by the need to quickly reach adult form and function, they improve their techniques through trial and error and observational learning (Whiten, 1989; Visalberghi and Fragaszy, 1990, 2002). As suggested by learning models, juvenile individuals gradually overcome the developmental constraints and initial differences from adult foragers (Marchetti & Price, 1989). Larger juveniles and adults, with better physical abilities, can handle more challenging prey (Gibson, 1986; Fragaszy and Boinski, 1995). Juveniles show similar diet to adults, but their foraging is shaped by risk aversion and limited motor skills (MacKinnon, K.C. 2006). Young ruminants, for example, adapt by selecting nutrient-rich plant parts, enhancing digestion and absorption despite higher time and energy costs (Van Soest, 1982). Juvenile primates, with smaller body sizes and higher metabolic needs per unit body weight, prefer high-quality foods like animal matter, rich in protein and energy, and avoid hard-to-digest fibrous foods (Iwamoto, 1982; Kleiber 1987; Nakayama et al., 1999).

Urban adaptation is likely to influence the foraging patterns in different life stages or urban adapters. In FRDs, pups begin weaning by the 7th week, with nursing ceasing entirely by the 13th week of age as they transition to independent foraging (Paul & Bhadra, 2017). At this stage, young animals are faced with foraging inexperience while catering for higher energy needs and faster digestion rates compared to adults (Arnold and Hill, 1972; Custance et al., 1999; Marchetti & Price, 1989). Juveniles in FRDs have been shown to be less choosy in their eating habits, which might lead to unpleasant experiences and even the risk of food poisoning.

In India, free-ranging dogs (FRDs) live in various human environments showing a strong preference for meat even when consuming an omnivorous diet (Sen Majumder et al., 2016; Bhadra et al., 2016b; Sarkar et al., 2019). They are key players in the scavenging community in human-dominated habitats, and have adapted their diet to include a variety of foods, including meat and vegetable scraps (Bhadra et al., 2016). They frequently choose food of a lower quality over higher-quality options (Vicars et al., 2014; Cameron et al., 2021) to avoid rivalry (Sarkar et al., 2023). Utilizing their keen sense of smell to identify protein-rich food is credited with their success in scavenging (Bhadra et al., 2016; Sarkar et al., 2019). Dogs can selectively locate and choose preferred food from garbage using olfactory signals (Sarkar et al., 2019). In experiments, they distinguish between different quantities of their preferred food, primarily based on olfactory cues (Banerjee & Bhadra, 2019). While scavenging from garbage,

FRDs often encounter various non-preferred and inedible items mixed with food that they prefer to eat, like meat and fish bones, rice, bread, some fruits and vegetables (Sarkar et al., 2019, Butler et al., 2018). One of the items that they encounter quite often in garbage is lemon, as Indian diet includes lemon quite frequently, both in the raw form as well as in cooked food. Dogs that are kept as pets and rely on their owners for food (Bhattacharjee et al., 2017), are often reported to dislike citrus flavours, including lemon (Woodford & Griffith, 2012). Previous studies show clear avoidance of adult FRDs towards whole lemon, and lemon juice (Pal et al., 2024). Lemon pulp and rind are less avoided than lemon juice and adult FRDs preferentially scavenge from less concentrated lemon juice environments over more concentrated ones (Pal et al., 2024). The food preferences and discrimination abilities of juvenile FRDs have not been adequately explored. It is important to understand whether juveniles exhibit similar preferences and avoidance patterns, particularly in relation to the acidity and concentration of available food sources. Juvenile FRDs are in a critical developmental stage and their scavenging behaviour may be influenced by various factors, such as learning through social interactions with adults and physiological changes. Therefore, investigating the food preferences of juvenile FRDs can provide valuable insights about developmental adaptive processes in the foraging behaviour of FRDs.

This study aimed to explore how FRDs of different ages (juvenile vs adult) make decisions in response to unpleasant stimuli (aversive sour liquid). The study examined how the dogs' food choices were influenced by the concentration and acidity (pH) of the unpleasant environment. To achieve this, we conducted food choice tests where juvenile and adult free-ranging dogs were presented with palatable food with varying concentrations of aversive liquid environments.

## 2. Materials and methods:

### 2.1 Study sites

The study was conducted in 13 different locations within the Nadia district (22.9747° N, 88.4337° E) of West Bengal, India (Fig 1). The experiments were conducted in two time slots: 06:00 -12:00h and 15:00 - 20:00h at the randomly selected sites.

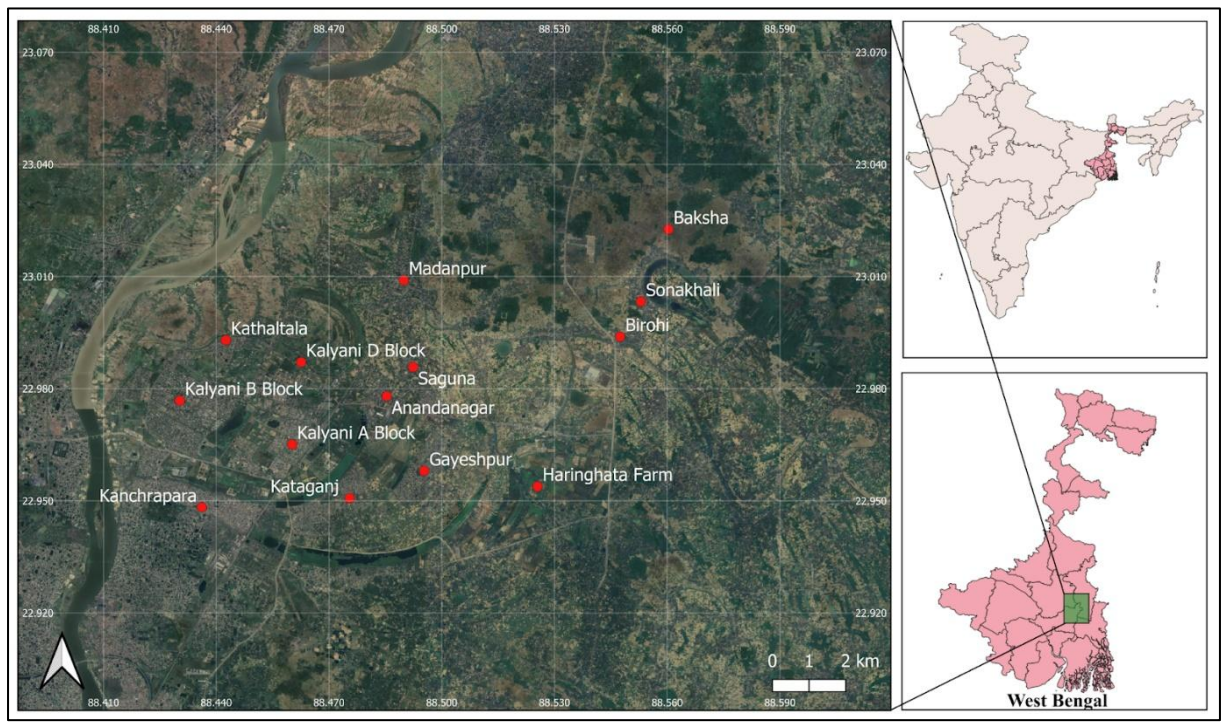

Fig.1 Study area map for experimental trials on free-ranging dogs (FRDs).

### 2.2 Subjects

The study was conducted on 73 juvenile FRDs that appeared to be healthy, giving no indications of illness or injury. The gender of each FRD was recorded during the investigation by observing their genitalia. Preference was given to single juvenile dogs, and interference from other gathering individuals or groups were avoided as best as possible during the test. The experiments were conducted in the pup-rearing and pre-mating seasons (winter, spring and summer), when juveniles are typically available in the population. Juveniles were identified by their smaller body size and underdeveloped reproductive features, i.e., the males had undescended testicles, and females had immature and underdeveloped mammary glands. Only those individuals that participated in the experiments on being presented with the set-up were used for the trials.

**2.3 Methodology of the choice test**

**2.3.1 Three bowl, three concentration choice test on juveniles.**

Fresh lemon juice was squeezed and diluted using distilled water to create solutions with concentrations of 50%, 33.3%, and 25% volume by volume, labelled as solutions 1, 2, and 3, respectively. In each trial, three paper bowls containing these lemon juice solutions, along with a fresh chicken piece weighing approximately 15 grams were spaced equally on a cardboard piece. The placement order of the bowls was randomized for each trial, and clearly noted down at the beginning of the experiment. The experiment involved placing the setup on the ground, about 1-1.5 meters in front of each dog, and video recording the behaviours of the dogs for 180 seconds. After placing the set-up on the ground, the experimenter took a few steps backwards a looked straight ahead, avoiding eye contact with the focal dog. Precautions such as random selection of dogs and minimizing disturbances were carried out to reduce biases. After each trial, the bowls were replaced to ensure consistency and prevent cross-contamination. Experiments were conducted under similar environmental conditions to minimize variability.

The methodology for the adult FRDs has been previously detailed in Pal et al., 2024. The adult study involved similar experimental setups and conditions, allowing for a comparative study between juvenile and adult FRDs in response to food with different concentrations of lemon juice in their scavenging behaviour.

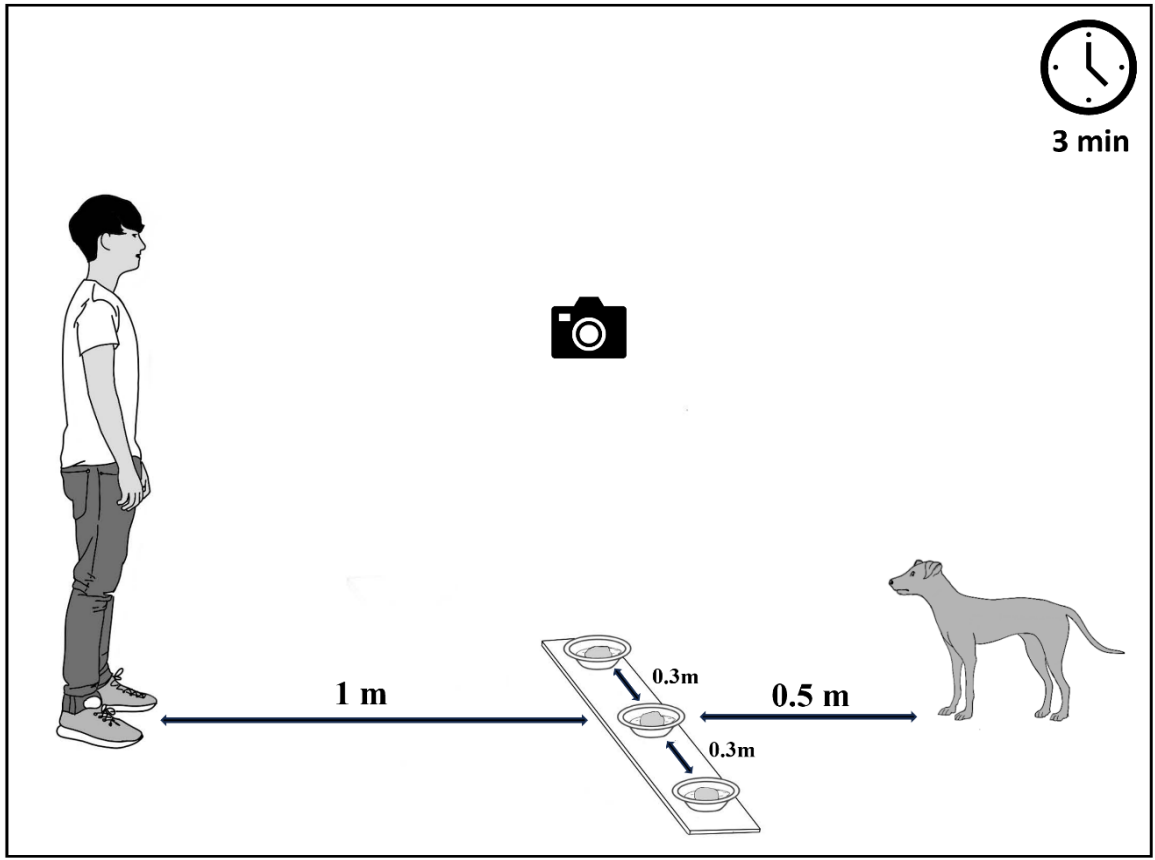

Fig. Experimental design for three choice tests on juvenile dogs.

**2.3.2 pH assessment of lemon juice solutions.**

Fresh lemon juice was extracted from washed and dried lemons. The juice was filtered through a fine mesh sieve to remove pulp and seeds, ensuring a consistent concentration of juice in the solutions. Three solutions consisting of 25%, 33.33% and 50% concentration of lemon juice were prepared by mixing 100% freshly extracted juice with three parts, two parts and equal parts of distilled water respectively. Their pH values were calculated using a pH meter. For calibrating, predefined solutions of pH 7 and pH 4 were used and the electrodes were carefully rinsed and blotted after each sampling. Each concentration of lemon juice was sampled 15 times.

**3. Behavioural analysis**

Specific behaviours displayed by the focal FRDs were recorded from the videos by two observers. Inter-rater reliability was estimated prior to data analysis. A detailed ethogram was prepared listing all strategic behaviours used to handle the food (Table 1). Latency was defined as the time spent between placement of the bowl to the first sniff response by the dog. A sniff was recorded as a valid part of interaction when the nose was brought within a distance of approximately 10 cm from the bowl. A complete act of eating involved engulfing all parts of the chicken piece within the mouth and completely consuming it. The time investment on each bowl was checked and decoded in seconds.

| Behaviour | Definition |
|---|---|
| Sniffing | The dog brings its nose within 0.1 m of the bowl to investigate the scent of the contents. |
| Licking & Vigorous licking | Involves the active use of the tongue to taste or clean the food or liquid present in the bowl. Vigorous licking refers to instances where dogs lick more than 3-4 times consecutively. |
| Strategizing | Observed when dogs purposefully refrained from immediately consuming a food item after sniffing or licking it, instead manipulated it with various body parts like tongue, foreleg, muzzle etc. |
| Eating | Actively taking food into the mouth, resulting in complete absorption or swallowing. |

Table. 1 Ethogram: A list of behaviours coded from experiment videos and their definitions.

**Statistical analysis**

We performed the Mann-Whitney U Test to compare the interaction times between adult and juvenile dogs for each concentration. The Kruskal-Wallis H Test was used to compare the time investment data across different concentrations. The Chi-Square Test was applied to compare behavioural frequencies (sniffing, licking, strategizing, and eating) across different concentrations and between adult and juvenile dogs at each concentration. For the pH level data, we used the Kruskal-Wallis Test, followed by post-hoc Dunn's tests for pairwise comparisons across concentrations. To analyse time investment data, we converted the time spent at each of the three bowls into proportions by dividing each by the total time investment. We applied beta regression (using the betareg package) to examine the influence of concentration, latency, eating events, and gender on the first interaction time and total time investment. Additionally, we used beta generalized linear mixed models (GLMMs) via the glmmTMB package to assess the effect of predictors on the proportion of first interaction time, incorporating random effects. The lme4 package was employed to analyse the impact of predictors on the probability of eating outcomes and strategizing behaviour. Model performance was evaluated using the performance package.

For Markov Chain analysis, we extracted ordered event sequences from cleaned data, identified state transitions, and computed transition probabilities based on observed frequencies. The transition probability matrix was constructed by normalizing transition counts so that each row summed to 1. The state transitions and their probabilities were visualized as a directed weighted graph using NetworkX (Python) (Hagberg et al., 2008) and DiagrammeR (R). We estimated steady-state probabilities using the steadyStates() function from the Markovchain R package, representing the long-term proportion of time spent in each state. Sankey diagrams were used to visualize behavioural transitions and frequencies in juvenile and adult free-ranging dogs, highlighting differences in lemon avoidance strategies. All statistical analyses were conducted using R Statistical Software (v4.2.2; R Core Team, 2022).

## Results
### 1. Frequency of different events in juvenile FRDs.

After placement of the set-up, juvenile dogs approached and sniffed the food provided in all successful trials. The juvenile dogs sniffed and licked all three choices equally ($\chi^2 = 3.00$, df = 2, $p > 0.05$ for sniffing; $\chi^2 = 1.30$, df = 2, $p > 0.05$ for licking). Further, they did not show any preference for strategizing ($\chi^2 = 0.58$, df = 2, $p > 0.05$) or eating first ($\chi^2 = 1.51$, df = 2, $p > 0.05$) with respect to the concentration of lemon juice. These results indicate that the concentration of lemon juice did not significantly affect the initial foraging behaviours of juvenile dogs.

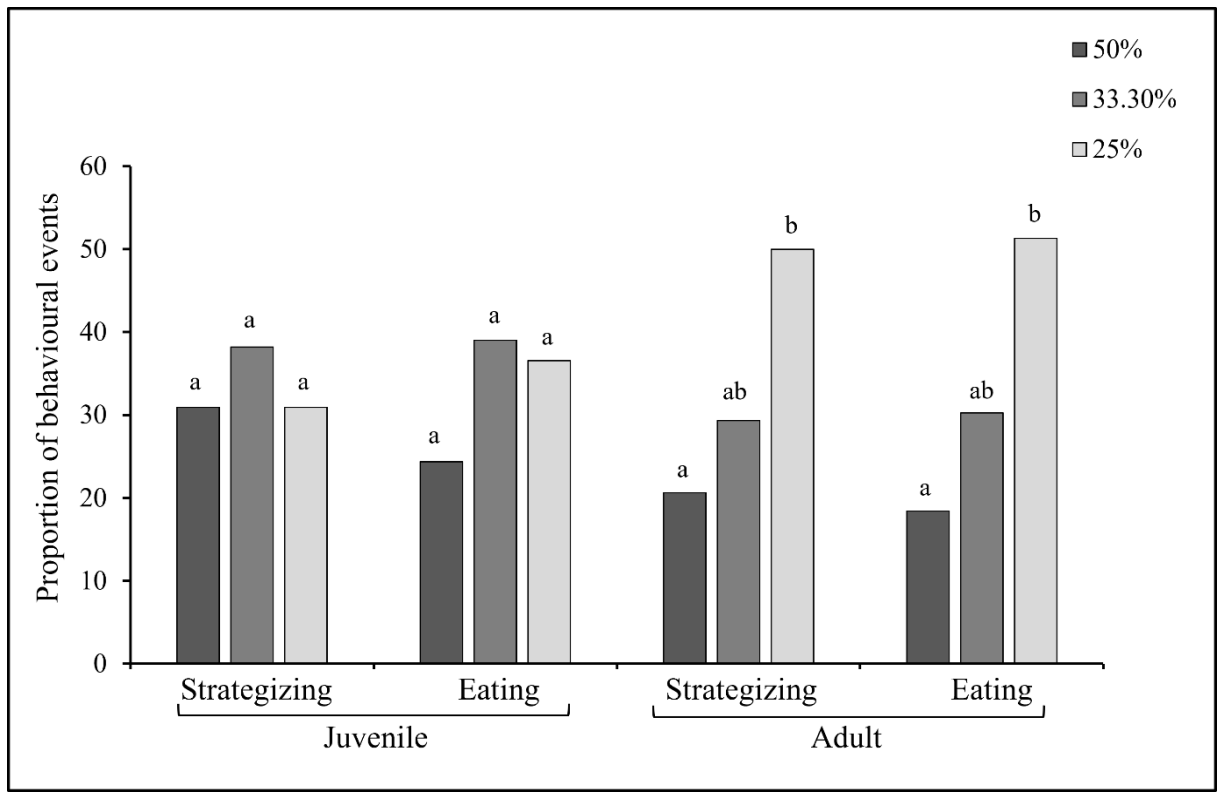

Fig 2: Percentage of first choices per event across different concentrations of lemon juice for adult and juvenile FRDs. Different letters (a, b) indicate significant differences. Groups sharing the same letter are not statistically significant.

In our previous study (Pal et al., 2024), we investigated the foraging behaviours of adult dogs in response to varying food concentrations. The current research focuses on juvenile dogs, allowing for a comparative analysis of behavioural responses between these two age groups. Juvenile dogs demonstrated consistent behaviours across different food concentrations, showing no significant differences in sniffing, licking, strategy-making, or eating behaviours. In contrast, adult dogs exhibited significant variations in their foraging strategies. Specifically, adult dogs displayed a marked difference in their initial strategy-making ($p < 0.05$) and total eating events ($p < 0.05$) when comparing the highest (50%) and lowest (25%) food concentrations. Notably, adult dogs were more likely to strategize and consume food from the lower concentration bowl (25%) compared to the higher concentration bowl (50%). These findings suggest that adult dogs exhibit a more selective and adaptive foraging behaviour based on food concentration, whereas juvenile dogs maintain a more uniform response regardless of concentration (Fig 2).

We investigated behavioural differences in the first occurrences in events between adult and juvenile dogs across three concentrations (50%, 33%, and 25%) using Chi-Square tests. No significant differences were observed in sniffing frequencies between adult and juvenile dogs at any concentration level. Specifically, at 50% ($\chi^2 = 0.609$, $p > 0.05$), 33% ($\chi^2 = 1.731$, $p > 0.05$), and 25% ($\chi^2 = 0.178$, $p > 0.05$) concentrations, the age groups displayed similar levels of sniffing. Similarly, licking behaviour was comparable between adult and juvenile dogs across all concentrations (50%: $\chi^2 = 1.38 \times 10^{-30}$, $p > 0.05$; 33%: $\chi^2 = 0.006$, $p > 0.05$; and 25%: $\chi^2 = 0.761$, $p > 0.05$). Strategizing behaviour showed no significant age-related differences at 50% ($\chi^2 = 0.434$, $p > 0.05$) and 33% ($\chi^2 = 0.078$, $p > 0.05$) concentrations. However, a significant difference emerged at the 25% concentration ($\chi^2 = 7.06$, $df = 1$, $p < 0.001$). This difference was accompanied by a moderate effect size (Cramér's V = 0.199), indicating a notable association between age and strategizing behaviour at this lower concentration. Adults showed higher level of strategizing as compared to the juveniles. Eating behaviour did not significantly differ between adult and juvenile dogs at 50% ($\chi^2 = 1.10 \times 10^{-30}$, $p > 0.05$) and 33% ($\chi^2 = 0$, $p > 0.05$) concentrations. However, a significant difference was observed at the 25% concentration ($\chi^2 = 4.85$, $df = 1$, $p < 0.05$), albeit with a small effect size (Cramér's V = 0.165). This suggests that at the 25% concentration, age influenced eating behaviour, and once again, adults ate at a higher frequency than the juveniles.

2. **Duration of scavenging activity in Free-Ranging Dogs**

The first interaction times of adult dogs with chicken bowls containing 50%, 33.3%, and 25% lemon juice concentrations differed significantly (Kruskal-Wallis test: $H = 6.54$, $df = 2$, $p < 0.05$). Post-hoc Dunn's test (without adjustment) revealed significant differences between the 50% and 25% concentrations ($p < 0.01$) and between the 33.3% and 25% concentrations ($p < 0.05$), indicating that adult dogs invested more time in their initial interactions with lower lemon juice concentrations.

In contrast, juvenile dogs showed no significant differences in their first interaction times across the concentrations (Kruskal-Wallis test: $H = 0.91$, $df = 2$, $p > 0.05$). Similarly, no significant differences were observed in the total interaction times of adult dogs (Kruskal-Wallis test: $H = 1.04$, $df = 2$, $p > 0.05$), consistent with the findings for juvenile dogs (Kruskal-Wallis test: $H = 0.46$, $df = 2$, $p > 0.05$).

We examined the differences in both first interaction time and total interaction time between adult and juvenile dogs for the three different concentration levels (50%, 33%, and 25%). The first interaction time at the 50% concentration (W=4071, p > 0.05), 33% concentration (W=3648, p > 0.05), or 25% concentration (W=4435.5, p > 0.05) and total interaction time at the 50% concentration (W=4029.5, p > 0.05), 33% concentration (W=3824, p > 0.05), or 25% concentration (W=4467.5, p > 0.05) were comparable between adults and juveniles.

3. **Factors influencing eating behaviour.**

To understand the factors influencing the likelihood of eating behaviour in free-ranging dogs, a generalized linear mixed model (GLMM) with a binomial distribution and logit link function was employed. The model considered age class, concentration (conc.), gender, and their interactions as fixed effects, while dog ID was included as a random effect to account for individual differences.

$$eat \sim age\_class + concentration* gender + (1|dog.ID)$$

The results indicate that juvenile dogs had a significantly lower likelihood of engaging in eating behaviour compared to adults (Estimate = -1.041, $p < 0.001$). Similarly, male dogs were less likely to exhibit eating behaviour than females (Estimate = -0.605, $p < 0.001$).

Additionally, higher lemon juice concentrations (33.3% and 50%) were associated with a reduced likelihood of eating compared to the 25% concentration (33.3%: Estimate = -0.702, $p < 0.001$; 50%: Estimate = -0.622, $p < 0.001$). A significant interaction between gender and concentration suggests that the effect of concentration on eating behaviour varied by gender. Specifically, male dogs were less deterred by higher concentrations than females, showing a greater likelihood of eating at 33.3% (Estimate = 0.521, $p < 0.001$) and 50% (Estimate = 0.375, $p < 0.001$) compared to the 25% concentration (Supplementary figures. 1).

4. **Factors influencing the strategy-making behaviour.**

To understand the factors influencing strategy-making behaviour in free-ranging dogs, a generalized linear mixed model (GLMM) with a binomial distribution and logit link function was employed.

$$\text{strategy} \sim \text{age\_class} + \text{concentration} + (1|\text{dog.ID})$$

The results indicate that age class had a significant effect on strategy-making behaviour, with juvenile dogs being significantly less likely to engage in strategy-making compared to adults (Estimate = -0.636, p = 0.014). This corroborates the results from the earlier analysis. In terms of concentration levels, the 50% concentration was significantly associated with a reduced likelihood of strategy-making behaviour compared to the reference concentration (25%) (Estimate = -0.6283, p = 0.0271). However, the 33.3% concentration did not have a statistically significant effect on strategy-making behaviour (Estimate = -0.4013, p = 0.1590) (Supplementary figures. 2).

Additionally, the random effect for dog ID revealed substantial variation in strategy-making behaviour among individual dogs (Variance = 3.06, Std. Dev. = 1.749).

### 5. Factors influencing the proportion of first interaction time.

A beta regression model was utilized to examine factors influencing the proportion of first interactions in free-ranging dogs.

$$\text{First interaction time} \sim \text{concentration} + (1|\text{dog.ID})$$

The results indicate that higher lemon juice concentrations (33.3% and 50%) significantly reduced the proportion of first interactions. Specifically, the 33.3% concentration (Estimate = -0.271, p = 0.008) and the 50% concentration (Estimate = -0.364, p = 0.001) were associated with a decreased likelihood of strategy-making behaviour.

The random effect for dog ID showed minimal variance in the proportion of first interactions among individual dogs (Variance = 4.51e-11, Std. Dev. = 6.72e-06).

Overall, these findings suggest that higher concentrations of lemon juice (33.3% and 50%) significantly reduce the likelihood of first interactions in free-ranging dogs.

### 6. Variation in pH levels across different lemon juice concentrations.

The pH levels across the three lemon juice concentrations (50%, 33.3%, and 25%) showed a significant difference, as indicated by the Kruskal-Wallis test ($\chi^2$ = 38.81, df = 2, p < 0.001). Post-hoc Dunn's tests confirmed significant pairwise differences between all concentration levels, with pH values differing significantly between 50% and 33.3% (Z = -3.11, p = 0.0009), 50% and 25% (Z = -6.23, p < 0.001), and 33.3% and 25% (Z = -3.11, p = 0.0009). These findings indicate a clear gradient in acidity, where pH increases as lemon juice concentration decreases. The boxplot further illustrates this trend, showing the highest pH at 25% concentration and the lowest at 50%, demonstrating the dilution effect on acidity (Fig 3).

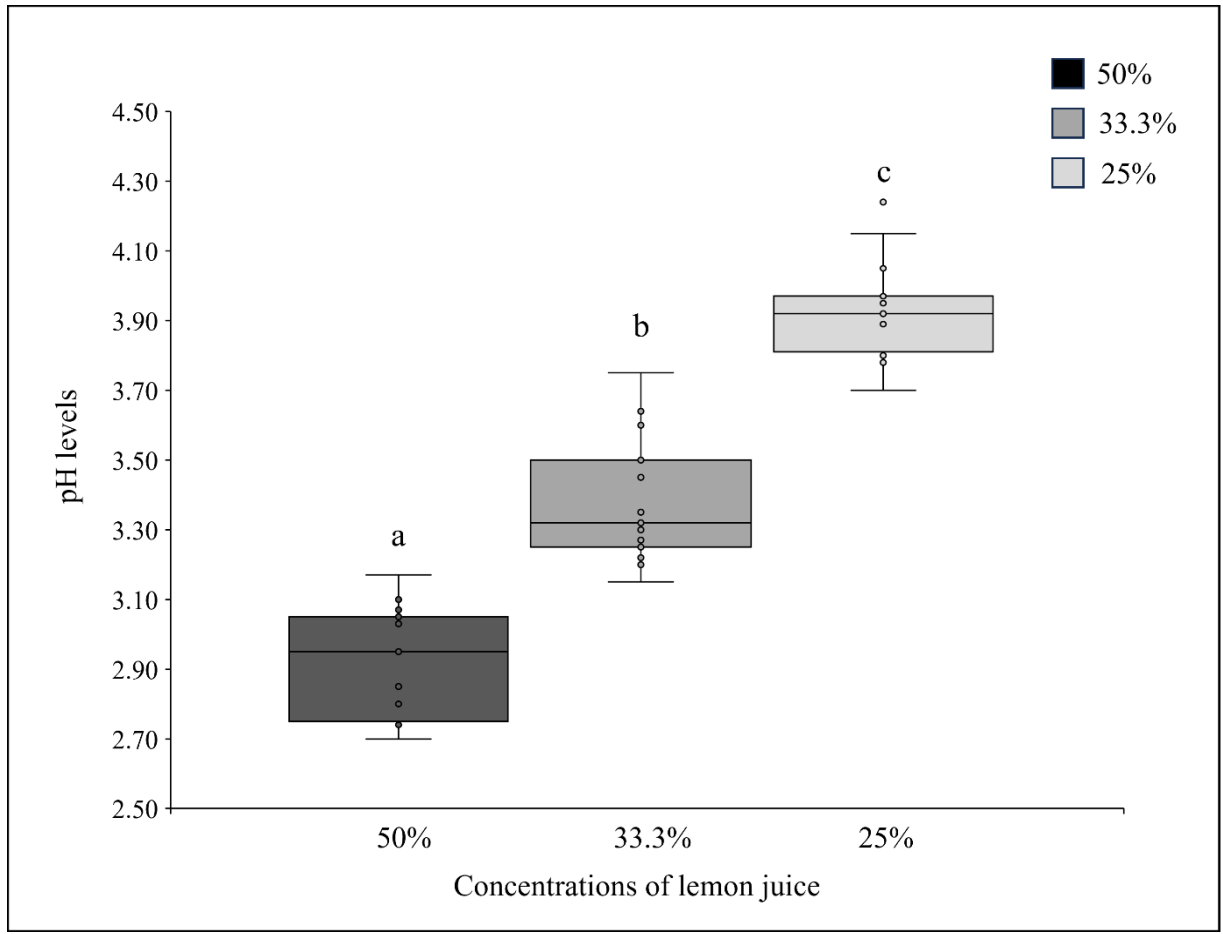

Fig 3: The mean pH of different concentrations of lemon juice shows significant difference. The Kruskal-Wallis H test: p = .001. Post-Hoc Dunn's test (α = 0.017): significant differences in mean ranks between 50%-33.3%, 33.3%-25%, and 50%-25%. Different letters (a, b, c) indicate significant differences. Different letters (e.g., a, b) indicate statistically significant differences between groups.

## 7. Markov chain analysis.

The Markov chain analysis revealed distinct state transition dynamics in juvenile and adult dogs across varying concentrations (50%, 33.3%, and 25%). For juvenile dogs, transitions from "Sniff" predominantly led to "Lick" (71%, 74%, and 70% across concentrations), while "Strategize" showed a higher likelihood of transitioning to "Eat" (68%, 67%, and 60%). Notably, "Lick" frequently transitioned back to "Strategize" (81%, 85%, and 78%), indicating a cyclical behaviour. In adult dogs, transitions from "Sniff" to "Lick" were even more pronounced (89%, 88%, and 87%), and "Strategize" overwhelmingly led to "Eat" (85%, 92%, and 91%). Similar to juveniles, adults exhibited a strong tendency to transition from "Lick" back to "Strategize" (88%, 92%, and 94%). Across all scenarios, "Eat" served as the terminal state, with minimal transitions out. These results highlight significant differences in behavioural patterns between juvenile and adult dogs, particularly in their decision-making processes and state persistence, while underscoring the consistent role of "Lick" as a recurring state in both groups.

In juvenile dogs, the transition from Strategize → Eat exhibited the highest probability across all concentrations, significantly surpassing Sniff → Eat ($Z = -0.19$, $p = 0.848$) and Lick → Eat ($Z = 0.11$, $p = 0.909$). The difference between Sniff → Eat and Lick → Eat was significant at 50% concentration ($Z = -0.05$, $p = 0.953$) and 33.3% concentration ($Z = -0.05$, $p = 0.952$) but not at 25% concentration ($Z = -0.06$, $p = 0.948$).

Similarly, in adult dogs, Strategize → Eat remained the predominant transition, with a significantly higher probability than both Sniff → Eat ($Z = -23.80$, $p < 0.001$) and Lick → Eat ($Z = 15.12$, $p < 0.001$). The difference between Sniff → Eat and Lick → Eat was significant at 50% concentration ($Z = -3.69$, $p < 0.001$) and 33.3% concentration ($Z = -2.42$, $p = 0.015$) but did not reach significance at 25% concentration ($Z = -1.94$, $p = 0.052$) (Fig 4).

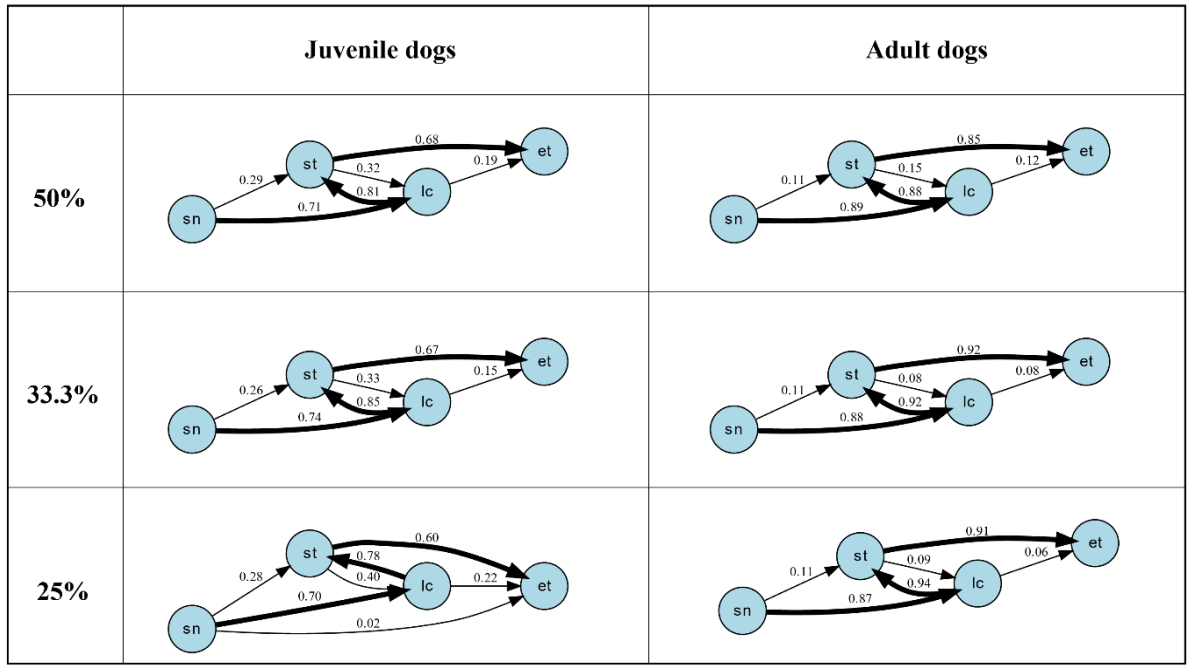

Fig 4. Markov chain transition diagrams illustrating behavioural transitions in juvenile and adult dogs across different concentration conditions. Each panel represents a different concentration level, with arrows indicating the probability of transitioning from one behaviour to another. Markov chain transition diagrams illustrating behavioural transitions in juvenile and adult dogs across different concentration conditions.

8. **Association between strategy use and eating behaviour in adult and juvenile FRDs.**

We examined the relationship between strategy use and eating behaviour in adult dogs. The results indicated a significant association ($\chi^2 = 182.0$, $p < 0.0001$). Pairwise comparisons revealed significant differences in the distribution of behaviour patterns across the groups: A significant difference was observed between dogs that showed strategy but did not eat and those that ate without showing a strategy ($\chi^2 = 38.1$, $p < 0.0001$), between dogs that showed strategy but did not eat and those that both showed strategy and ate ($\chi^2 = 58.3$, $p < 0.0001$), and between dogs that ate without strategy and those that showed strategy before eating ($\chi^2 = 151.0$, $p < 0.0001$).

In juvenile dogs, the Chi-square test revealed a significant relationship between strategy use and eating behaviour ($\chi^2 = 74.9$, $p < 0.0001$). Pairwise comparisons showed: A significant difference was found between dogs that showed strategy but did not eat and those that ate without showing strategy ($\chi^2 = 44.3$, $p < 0.0001$), between dogs that showed strategy but did not eat and those that both showed strategy and ate ($\chi^2 = 8.63$, $p = 0.0033$), and a highly significant difference was observed between dogs that ate without strategy and those that showed strategy before eating ($\chi^2 = 78.2$, $p < 0.0001$) (Fig 5).

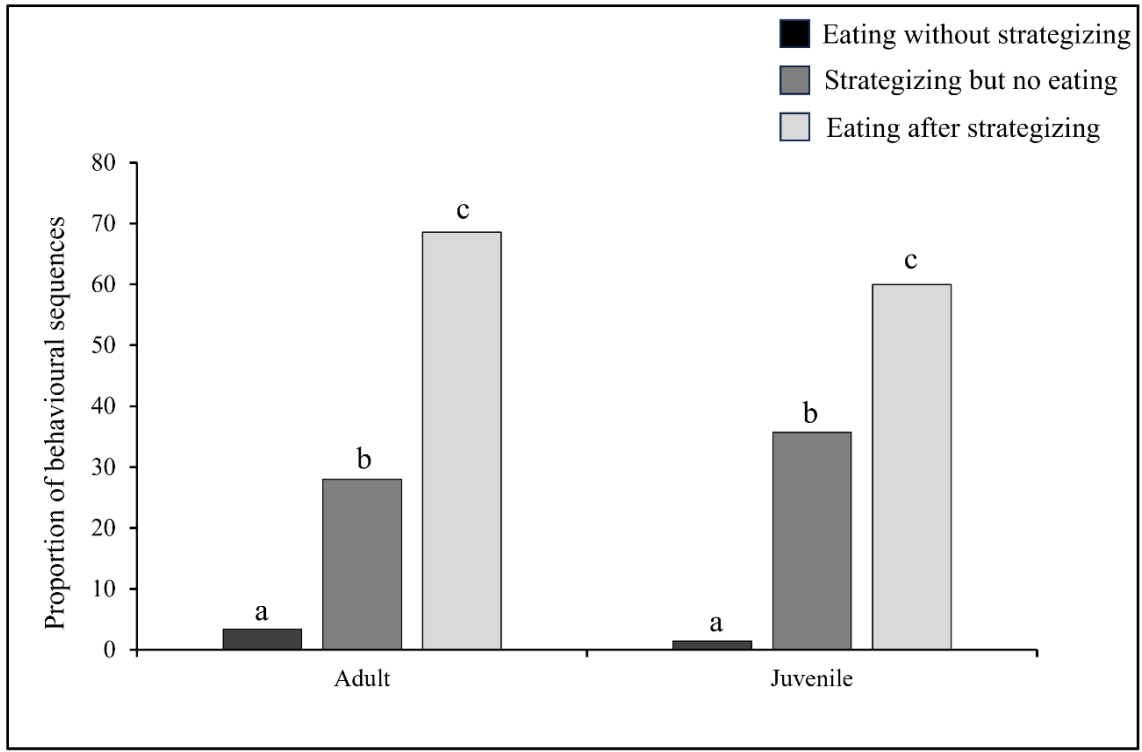

Fig 5. Comparison of strategize and eating behaviour in adult and juvenile Free-Ranging Dogs (FRDs). Different letters (e.g., a, b) indicate statistically significant differences between groups.

These findings suggest that strategy making is an important factor before eating behaviour by free-ranging dogs while they scavenge from acidic unpalatable environment, and that there may be developmental or environmental influences on whether dogs use a strategy before eating.

### 9. Behavioural flow and even transitions in adult and juvenile free-ranging dogs.

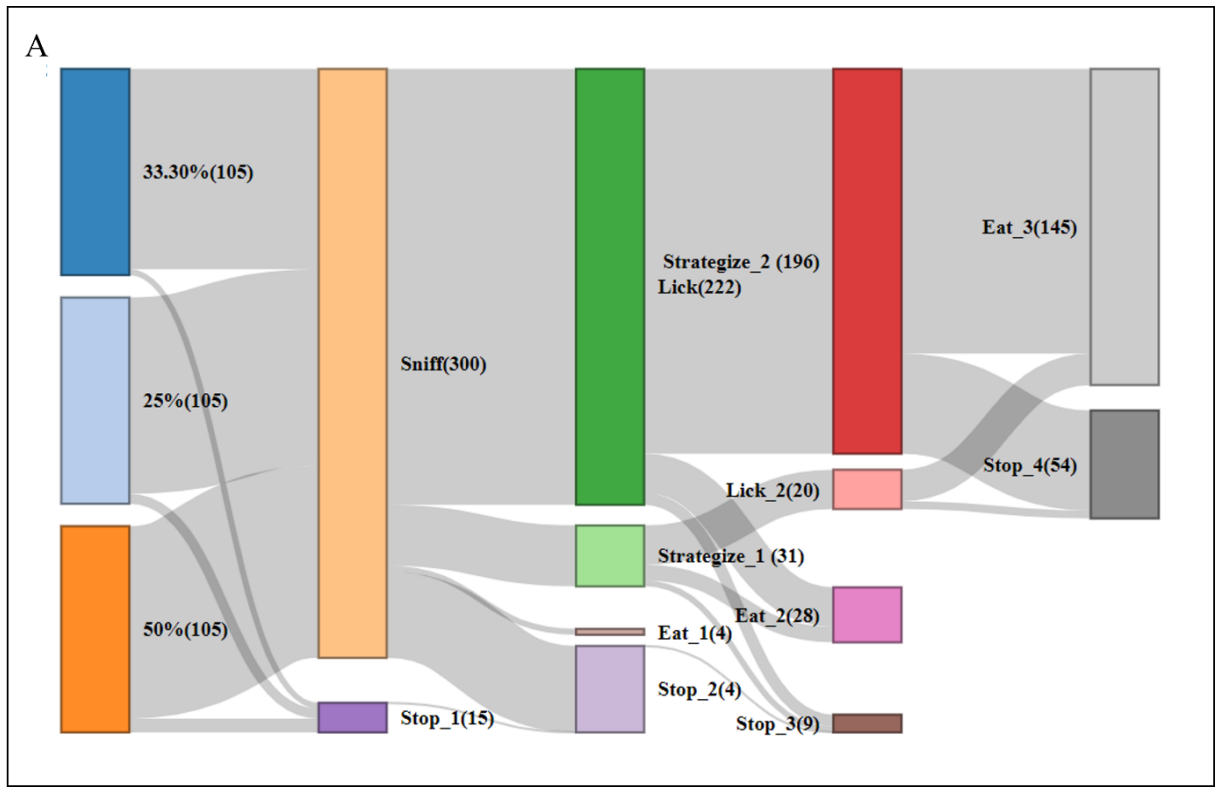

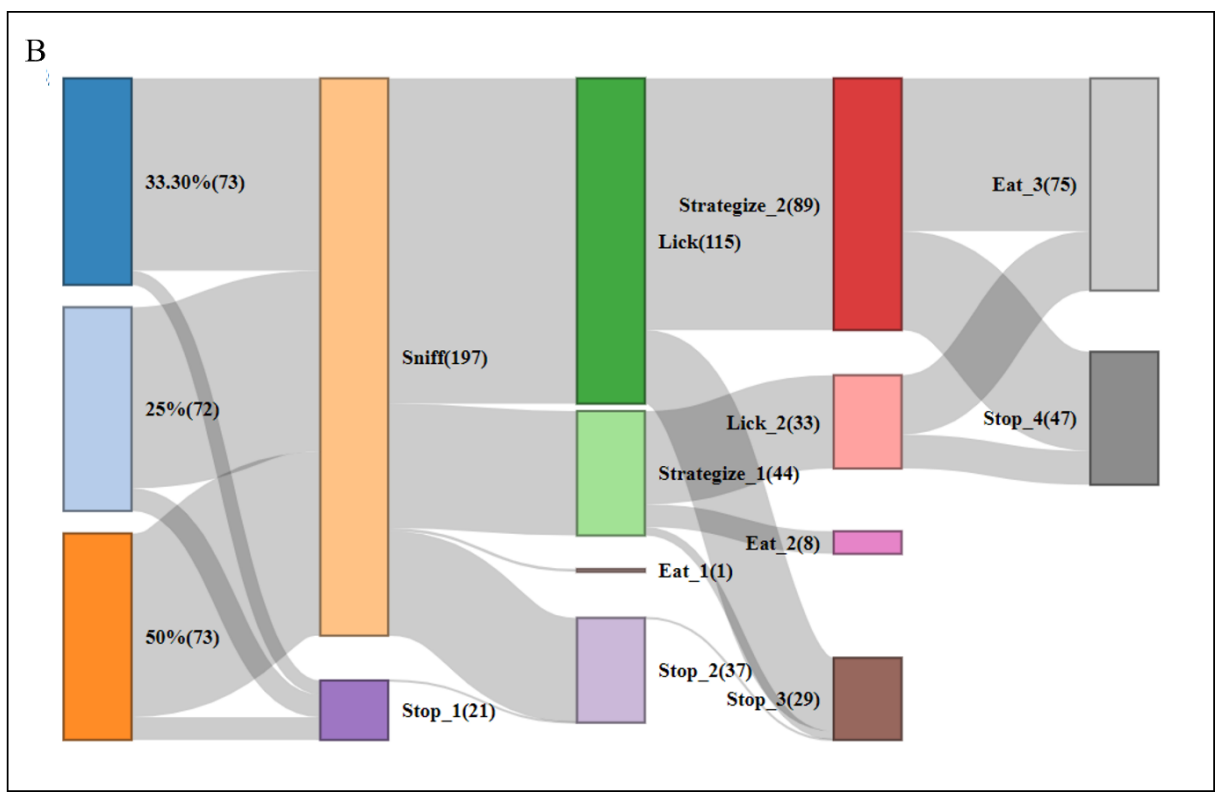

Fig 6. The Sankey diagrams depict the sequential flow of behavioural transitions in adult (Fig 6A) and juvenile (Fig 6B) free-ranging dogs, illustrating the progression from initial actions (e.g., sniffing, licking, strategizing) to final outcomes (e.g., eating or stopping). The diagrams

highlight differences in behavioural patterns between age groups, showing that juveniles exhibit more varied transitions before reaching the eating stage compared to adults. The numbers in the brackets represent the actual number of dogs that exhibited specific behavioural event or transition.

**Discussion**

This study investigated how juvenile and adult free-ranging dogs (FRDs) adapt their foraging strategies in response to aversive stimuli, specifically varying concentrations of lemon juice. Juvenile dogs responded uniformly across all concentrations, with no significant variation in any of the measured behaviours. This behavioural consistency suggests an underdeveloped ability to adapt to aversive cues, likely deriving from limited experience and immature decision-making skills. Since food preferences are heavily influenced by early learning, the juveniles tested were unable to distinguish between stimuli that were unpleasant during foraging. These results align with previous findings that juvenile animals often exhibit generalized, less discriminatory foraging patterns (Capretta & Bronstein, 1967; Capretta, 1969). In contrast, adult dogs showed a clear preference for lower acidity, displaying more strategic behaviours and exhibiting greater selectivity in their eating responses (Pal et al., 2024). Their refined strategies reflect accumulated experience and enhanced sensory discrimination towards unpalatable stimuli. Adults also invested more time during initial interactions with food at lower concentrations, supporting the idea that they assess risk before consumption, possibly guided by early sensory cues. Notably, the proportion of first interactions was significantly lower at higher concentrations, further underscoring a deterrence effect at the earliest stage of approach. According to Provenza and Malechek (1986), these patterns are consistent with developmental trends seen in other species, such as young goats, which need more time and expertise to forage efficiently. Similarly, prior studies on FRDs show that juveniles often consume indiscriminately, while adults avoid unpalatable items (Bhadra & Bhadra, 2014). Overall, our results highlight how experience and age influence adaptive scavenging tactics in dogs who roam freely and must navigate uncertain and occasionally unpleasant situations.

The concentration of lemon juice, age, and gender all had significant impacts on eating behaviour. Compared to adults, juveniles were less likely to consume, demonstrating how maturity and experience influence foraging decisions. This may arise from juveniles' inability to differentiate edible components from the aversive lemon environment, a limitation common across species navigating complex environments. Males were also less likely than females to consume, which may indicate that foraging behaviour is influenced by sex-based differences in risk tolerance or food preference.

Higher lemon concentrations (33.3% and 50%) reduced the likelihood of eating, reflecting an aversive sensory response. Notably, the interaction between gender and concentration depicted that male dogs were somewhat more tolerant of higher acidity which indicates sex-specific variation in sensory sensitivity or foraging resilience. These findings emphasize how environmental cues, physiological factors, and individual traits interact to influence food acceptance (Page & Jones, 2016; Sulikowski, 2017).

Strategizing during foraging seemed to depend both on the dogs' age and the concentration of lemon juice. Juvenile dogs were less likely to show strategic behaviour compared to adults, which fits with the idea that younger animals are still developing the cognitive skills needed for flexible decision-making. Similar developmental patterns are observed in species like the wandering albatross, where juveniles require time to learn optimal foraging tactics (Riotte-Lambert & Weimerskirch, 2013). Higher concentrations, particularly 50%, further reduced strategy-making behaviours, suggesting that stronger aversive cues suppress complex decision-making in favour of instinctive avoidance. Interestingly, some dogs just seemed naturally more inclined to strategize than others no matter what the situation was, suggesting that, like people, dogs can vary in how they approach problems or make decisions.

Higher concentrations significantly decreased initial engagement, according to the first interaction proportions beta regression analysis. This suggests that dogs utilize early sensory cues, likely olfactory or gustatory, to rapidly assess and avoid undesirable food. The minimal variance in individual-level responses indicates a shared sensory threshold across the sample, emphasizing the dominance of environmental signals in guiding initial food approach (Howery et al., 2000, 2014).

Collectively, these findings reveal the sophisticated interplay of development, sensory processing, and individual differences in shaping foraging responses. Adult dogs display greater adaptability, using learned strategies to mitigate risk, while juveniles exhibit a more uniform and less nuanced approach. The observed sex differences in eating behaviour further suggest intrinsic factors, possibly nutritional or hormonal, may modulate risk perception during feeding. Lemon juice concentration emerged as a central driver, consistently deterring interaction, strategizing, and eating at higher levels, thereby highlighting its efficacy as an aversive stimulus. Our study reveals the intricate interplay between age, gender, and environmental stimuli in shaping scavenging behaviour among free-ranging dogs. A particularly notable pattern emerged in juveniles: those that engaged in strategy-making were significantly more likely to eat, suggesting that even at an early developmental stage, cognitive planning can enhance foraging success in aversive contexts.

We have noticed that juveniles who strategized but did not eat differed significantly from those who ate without strategizing and from those who both strategized and ate. This indicates that while strategic behaviour is beneficial, its success depends on additional factors—such as sensory evaluation or perceived palatability. The difference between dogs that ate without strategizing and those that planned their actions highlights the functional value of preparatory behaviours, especially in unfamiliar or challenging environments.

These trends suggest that strategy-making in juveniles may reflect an emerging capacity for adaptive decision-making. Although adults are more adept due to greater experience, juveniles appear to be undergoing a learning phase, experimenting with and refining their behavioural responses. Similar learning curves have been observed in seabirds, where juveniles are initially inefficient foragers and gradually improve through trial and error (Wunderle, 1991). Supporting this, the Markov chain analysis revealed behaviour loops, such as from lick to strategize and then eat particularly noticeable in adults. These loops may indicate a behavioural evaluation cycle where sensory information is continuously assessed before consumption. Juveniles, though less consistent in such patterns, show early signs of this evaluative loop, reflecting developing decision-making structures. The observed gender differences in eating behaviour, with males being less likely to eat but also less deterred by higher acidity, point to sex-specific strategies, possibly shaped by differing risk thresholds or physiological needs.

Taken together, these findings suggest that decision-making in free-ranging dogs is shaped not only by immediate sensory cues but also by age, experience, and cognitive flexibility. Strategy-making, especially under aversive conditions, emerges as a key behavioural event for optimizing foraging outcomes. Over time, prolonged exposure to environmental challenges may further drive the development of novel foraging tactics. The emergence of adaptive scavenging techniques in these animals is probably influenced by the interaction of social learning, cognitive maturity, and environmental input. In the end, this study emphasizes how cognitive processes, such evaluative feedback and strategic planning, support successful foraging, thereby establishing the wider connection between ecological adaptability, survival, and decision-making in all species.

**Ethical Statement**

The study design did not violate the Animal Ethics regulations of the Government of India (Prevention of Cruelty to Animals Act 1960, Amendment 1982). The protocol for the experiment was approved by the IISER Kolkata Animal Ethics Committee.

**Conflict of interest statement**

All the authors have read and agree with this version of the manuscript. The authors declare no conflict of interest.

**Authors' contributions**

The study was conducted by TSP with the assistance of PD and SB. TSP, PD, and SB were responsible for data collection. The experiment was designed by TSP under the guidance of AB. PD performed the decoding, while TSP and SB carried out the analysis. Illustrations were created and edited by SB and TSP. The manuscript was written by TSP and SB, with guidance from AB. AB supervised the study and provided valuable input during the writing process. PD and SB contributed equally as joint second authors.

**Acknowledgements**

We thank Rohan Sarkar, Kushal Shaw and Dr. Udipta Chakraborti for their guidance and valuable suggestions. The authors also thank Imran Mondal for his help in the field for data collection. Additionally, we are grateful to all the members of the Dog Lab and the Behaviour and Ecology Lab (BEL) for their continuous support and help throughout the course of this study.

**Funding**

The research conducted by TSP was funded by the University Grant Commission. Additionally, this project received support from the Janki Ammal Grant and the Annual Research Fund provided by IISER Kolkata.

**Supplementary data**

Will be made available on publication of the manuscript after peer review.

**Supplementary information**

**Supplementary Table 1: Chi-Square Test Results for First Behavioural Events in Juvenile Dogs**

| Behaviour Event | Sample Size (n) | Chi-Square Statistic | p-value | Degrees of Freedom (df) | Method | Significance (p.signif) |
|---|---|---|---|---|---|---|
| **First Sniffing Events** | 3 | 0.3562 | 0.837 | 2 | Chi-square test | ns |
| **First Licking Events** | 3 | 1.3000 | 0.522 | 2 | Chi-square test | ns |
| **First Strategizing Events** | 3 | 0.5818 | 0.748 | 2 | Chi-square test | ns |
| **First Eating Events** | 3 | 1.5122 | 0.469 | 2 | Chi-square test | ns |

**Supplementary information 1.**

1. The transition matrices for juvenile and adult dogs at varying concentrations are summarized below, indicating the probabilities of transitioning between states.

1. **Juvenile Dogs (50% Concentration):**
   - From "Sniff" to "Strategize": 29%, "Sniff" to "Lick": 71%
   - From "Strategize" to "Lick": 32%, "Strategize" to "Eat": 68%
   - From "Lick" to "Strategize": 81%, "Lick" to "Eat": 19%
2. **Juvenile Dogs (33.3% Concentration):**
   - From "Sniff" to "Strategize": 26%, "Sniff" to "Lick": 74%
   - From "Strategize" to "Lick": 33%, "Strategize" to "Eat": 67%
   - From "Lick" to "Strategize": 85%, "Lick" to "Eat": 15%

3. **Juvenile Dogs (25% Concentration):**
   - From "Sniff" to "Strategize": 28%, "Sniff" to "Lick": 70%
   - From "Strategize" to "Lick": 40%, "Strategize" to "Eat": 60%
   - From "Lick" to "Strategize": 78%, "Lick" to "Eat": 22%
4. **Adult Dogs (50% Concentration):**
   - From "Sniff" to "Strategize": 11%, "Sniff" to "Lick": 89%
   - From "Strategize" to "Lick": 15%, "Strategize" to "Eat": 85%
   - From "Lick" to "Strategize": 88%, "Lick" to "Eat": 12%
5. **Adult Dogs (33.3% Concentration):**
   - From "Sniff" to "Strategize": 11%, "Sniff" to "Lick": 88%
   - From "Strategize" to "Lick": 8%, "Strategize" to "Eat": 92%
   - From "Lick" to "Strategize": 92%, "Lick" to "Eat": 8%
6. **Adult Dogs (25% Concentration):**
   - From "Sniff" to "Strategize": 11%, "Sniff" to "Lick": 87%
   - From "Strategize" to "Lick": 9%, "Strategize" to "Eat": 91%
   - From "Lick" to "Strategize": 94%, "Lick" to "Eat": 6%

**Supplementary table 2: Chi-Square Test Results for Strategy Usage Before Eating in Adults and Juveniles**

| Group | Strategy Used | Frequency | Chi-Square Statistic | df | p-value | p.signif |
|---|---|---|---|---|---|---|
| **Adults** | "10" (Show but no eat) | 56 | 182.0 | 2 | 3.47e-40 | **** |
| | "01" (Eat without strategy showing) | 7 | | | | |
| | "11" (Show then eat) | 171 | | | | |
| **Juveniles** | "10" (Show but no eat) | 50 | 74.9 | 2 | 5.49e-17 | **** |
| | "01" (Eat without strategy showing) | 2 | | | | |

|  | "11" (Show then eat) | 84 |  |  |  |  |

**Supplementary table 3: Pairwise Chi-Square Post Hoc Comparisons for Strategy Usage**

| Group | Comparison | Chi-Square Statistic | df | p-value | p.adj | p.adj.signif |
|---|---|---|---|---|---|---|
| **Adults** | "10" vs "01" | 38.1 | 1 | 6.68e-10 | 6.68e-10 | **** |
|  | "10" vs "11" | 58.3 | 1 | 2.3e-14 | 4.6e-14 | **** |
|  | "01" vs "11" | 151 | 1 | 9.96e-35 | 2.99e-34 | **** |
| **Juveniles** | "10" vs "01" | 44.3 | 1 | 2.81e-11 | 5.62e-11 | **** |
|  | "10" vs "11" | 8.63 | 1 | 3.31e-3 | 3.31e-3 | ** |
|  | "01" vs "11" | 78.2 | 1 | 9.38e-19 | 2.81e-18 | **** |

**Supplementary Table 4: Kruskal-Wallis and Dunn's Test Results for First Interaction Time in Juvenile Dogs**

**Kruskal-Wallis Test Results**

| Test | Chi-Square Statistic | df | p-value |
|---|---|---|---|
| **Kruskal-Wallis Test (Time ~ Concentration)** | 0.4576 | 2 | 0.7955 |

**Supplementary Table 5: Chi-Square Test Results for Side Preference and Gender Variation in Focal Dogs**

| Test | Chi-Square Statistic | df | p-value | Significance | Conclusion |
|---|---|---|---|---|---|
| **Side Preference (Left/Middle/Right)** | 1.30 | 2 | 0.522 | ns (not significant) | No significant difference in side preference. No pairwise chi-square test performed. |
| **Gender Variation in Side Preference** | 0.1233 | 1 | 0.7251 | ns (not significant) | No significant gender-based difference. No pairwise chi-square test performed. |

## Supplementary Material: Statistical Model Results

### 1. Generalized Linear Mixed Model (GLMM) - Binomial (Logit)

**Formula:**

*et ~ Proportion_1st_interaction + age_class + (1 | dog.ID)*

**Model Fit Statistics:**

- **AIC:** 609.8
- **BIC:** 626.6
- **Log-Likelihood:** -300.9
- **Deviance:** 601.8
- **Residual Degrees of Freedom:** 492

**Random Effects:**

- **dog.ID (Intercept):** Variance = 3.271, Std. Dev. = 1.809
- **Number of observations:** 496
- **Number of groups (dog.ID):** 122

**Fixed Effects:**

| Predictor | Estimate | Std. Error | z-value | p-value | Significance |
|---|---|---|---|---|---|
| (Intercept) | -0.1186 | 0.2903 | -0.408 | 0.682974 | |
| Proportion_1st_interaction | 1.7891 | 0.5312 | 3.368 | 0.000757 | *** |
| age_classjuvenile | -0.8766 | 0.2669 | -3.284 | 0.001024 | ** |

### 2. GLMM - Binomial (Logit)

**Formula:**

*et ~ age_class + conc. * gender + (1 | dog.ID)*

**Model Fit Statistics:**

- **AIC:** 672.1
- **BIC:** 706.3
- **Log-Likelihood:** -328.0
- **Deviance:** 656.1
- **Residual Degrees of Freedom:** 526

**Random Effects:**

- **dog.ID (Intercept):** Variance = 2.68, Std. Dev. = 1.637
- **Number of observations:** 534
- **Number of groups (dog.ID):** 122

**Fixed Effects:**

| Predictor | Estimate | Std. Error | z-value | p-value | Significance |
|---|---|---|---|---|---|
| (Intercept) | 1.0165 | 0.0011 | 915.0 | <2e-16 | *** |
| age_classjuvenile | -1.0410 | 0.0011 | -936.5 | <2e-16 | *** |
| conc.33(B) | -0.7024 | 0.0011 | -632.1 | <2e-16 | *** |
| conc.50(A) | -0.6221 | 0.0011 | -559.9 | <2e-16 | *** |
| genderM | -0.6054 | 0.0011 | -544.9 | <2e-16 | *** |
| conc.33(B):genderM | 0.5209 | 0.0011 | 468.8 | <2e-16 | *** |
| conc.50(A):genderM | 0.3751 | 0.0011 | 337.6 | <2e-16 | *** |

**Notes:**

- The model failed to converge (max|grad| = 0.0637, tol = 0.002).
- Consider rescaling variables due to near-unidentifiability.

### 3. GLMM - Binomial (Logit)

**Formula:**

*st ~ age_class + conc. + (1 | dog.ID)*

**Model Fit Statistics:**

- **AIC:** 605.4
- **BIC:** 626.8
- **Log-Likelihood:** -297.7
- **Deviance:** 595.4
- **Residual Degrees of Freedom:** 529

**Random Effects:**

- **dog.ID (Intercept):** Variance = 3.06, Std. Dev. = 1.749
- **Number of observations:** 534
- **Number of groups (dog.ID):** 122

**Fixed Effects:**

| Predictor | Estimate | Std. Error | z-value | p-value | Significance |
|---|---|---|---|---|---|
| (Intercept) | 1.8645 | 0.3182 | 5.859 | 4.66e-09 | *** |
| age_classjuvenile | -0.6363 | 0.2576 | -2.471 | 0.0135 | * |
| conc.33(B) | -0.4013 | 0.2849 | -1.408 | 0.1590 | |
| conc.50(A) | -0.6283 | 0.2842 | -2.211 | 0.0271 | * |

### 4. Generalized Linear Mixed Model (GLMM) - Beta (Logit)

**Formula:**

*Proportion_1st_interaction ~ conc. + (1 | dog.ID)*

**Model Fit Statistics:**

- **AIC:** -168.1
- **BIC:** -147.1
- **Log-Likelihood:** 89.0
- **Deviance:** -178.1

- **Residual Degrees of Freedom:** 491

**Random Effects:**

- **dog.ID (Intercept):** Variance = 4.512e-11, Std. Dev. = 6.717e-06
- **Number of observations:** 496
- **Number of groups (dog.ID):** 122

**Fixed Effects:**

| Predictor | Estimate | Std. Error | z-value | p-value | Significance |
|---|---|---|---|---|---|
| (Intercept) | -0.3581 | 0.0729 | -4.912 | 9.00e-07 | *** |
| conc.33(B) | -0.2714 | 0.1026 | -2.646 | 0.0081 | ** |
| conc.50(A) | -0.3642 | 0.1042 | -3.496 | 0.0005 | *** |

**Supplementary table: 6**

| Behaviour | Concentration | X-squared | df | p-value | Cramér's V |
|---|---|---|---|---|---|
| Sniffing | 50% | 0.60886 | 1 | 0.4352 | 0.05848577 |
| Sniffing | 33% | 1.7313 | 1 | 0.1882 | 0.09862335 |
| Sniffing | 25% | 0.17803 | 1 | 0.6731 | 0.03162514 |
| Licking | 50% | 1.38E-30 | 1 | 1 | 8.80E-17 |
| Licking | 33% | 0.0061833 | 1 | 0.9373 | 0.005893866 |
| Licking | 25% | 0.7611 | 1 | 0.383 | 0.06538985 |
| Strategizing | 50% | 0.43374 | 1 | 0.5102 | 0.04936362 |
| Strategizing | 33% | 0.078245 | 1 | 0.7797 | 0.0209662 |
| Strategizing | 25% | 7.0589 | 1 | 0.007887 | 0.1991401 |
| Eating | 50% | 1.10E-30 | 1 | 1 | 7.86E-17 |
| Eating | 33% | 0 | 1 | 1 | 0 |
| Eating | 25% | 4.8536 | 1 | 0.02759 | 0.1651281 |

**Supplementary table: 7**

| Concentration | Variable | W | p-value | Cliff's Delta (Estimate) |
|---|---|---|---|---|
| 50% | First Interaction Time | 4071 | 0.48 | 0.06223092 |
| 33% | First Interaction Time | 3648 | 0.5849 | -0.0481409 |
| 25% | First Interaction Time | 4435.5 | 0.07432 | 0.1573386 |
| 50% | Total Interaction Time | 4029.5 | 0.5609 | 0.05140248 |
| 33% | Total Interaction Time | 3824 | 0.9811 | -0.002217873 |
| 25% | Total Interaction Time | 4467.5 | 0.06047 | 0.1656882 |

**Supplementary figures. 1**

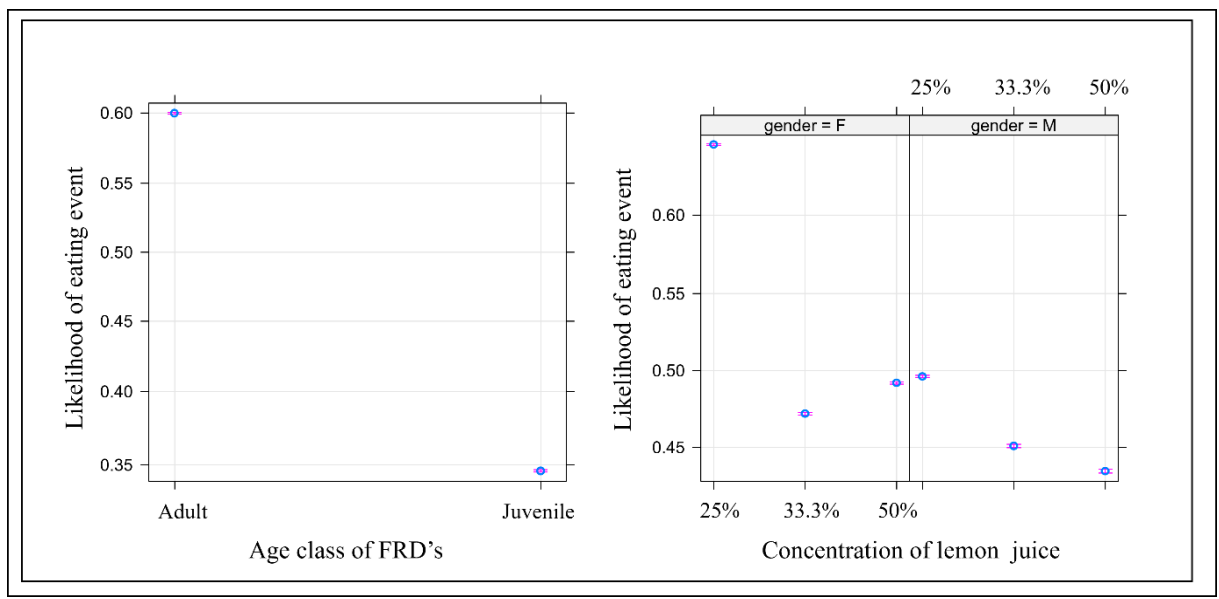

Fig. Effects of age, sex, and lemon juice concentration on the likelihood of eating behaviour in free-ranging dogs (FRDs).

**Supplementary figures. 2**

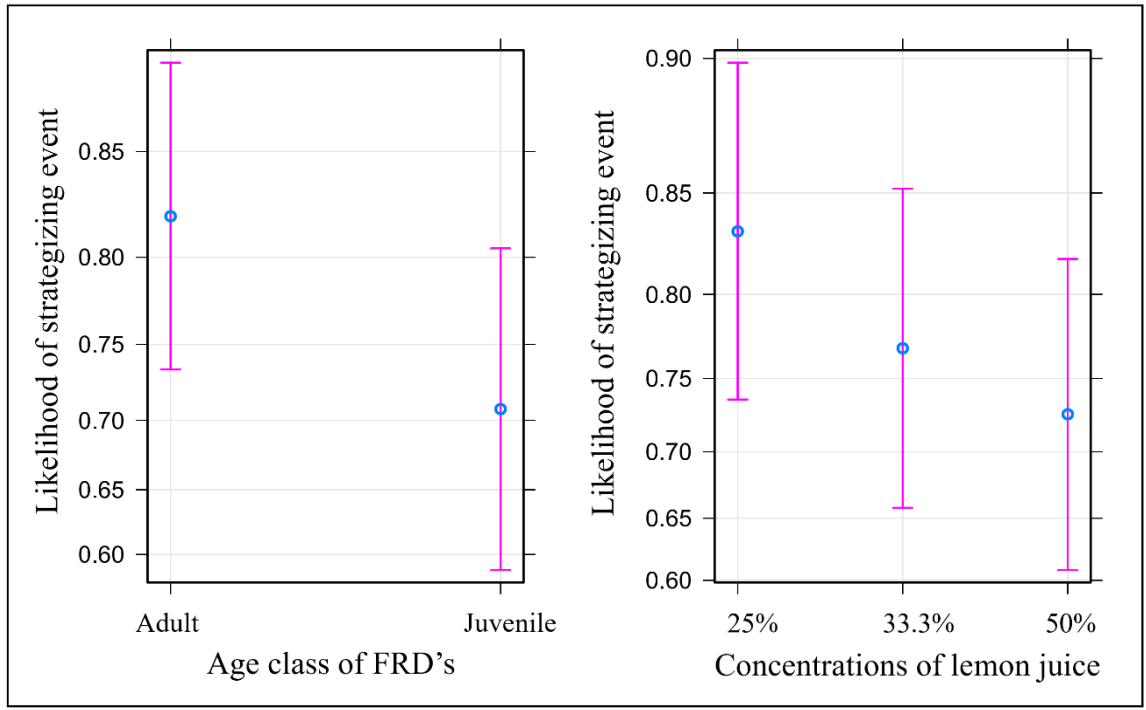

Fig. Influence of age and lemon juice concentration on the likelihood of strategizing behaviour in free-ranging dogs (FRDs).

------------